\documentclass[preprint,12pt]{article}

\usepackage{PRIMEarxiv}

\usepackage[utf8]{inputenc} % allow utf-8 input
\usepackage[T1]{fontenc}    % use 8-bit T1 fonts
\usepackage{hyperref}       % hyperlinks
\usepackage{url}            % simple URL typesetting
\usepackage{booktabs}       % professional-quality tables
\usepackage{amsfonts}       % blackboard math symbols
\usepackage{nicefrac}       % compact symbols for 1/2, etc.
\usepackage{microtype}      % microtypography
\usepackage{lipsum}
\usepackage{enumitem}
\usepackage{tabularx}
\usepackage{setspace}
\usepackage{fancyhdr}       % header
\usepackage{graphicx}       % graphics
\graphicspath{{media/}}     % organize your images and other figures under media/ folder

\title{An optimal replacement policy under variable shocks and different patterns of self-healing
}

\author{
  Debolina Chatterjee \\
  Department of Biostatistics and Health Data Science \\
  Indiana University School of Medicine (IUSM) \\
  Indianapolis, USA\\
  \texttt{dchatter@iu.edu} \\
   \And
  Jyotirmoy Sarkar \\
  Department of Mathematical Sciences \\
  Indiana University Indianapolis (IUI) \\
  Indianapolis, USA\\
  \texttt{jsarkar@iupui.edu} \\
}

\begin{document}
\maketitle

\begin{abstract}
We study a system that experiences damaging external shocks at stochastic intervals, continuous degradation, and self-healing. The motivation for such a system comes from real-life applications based on micro-electro-mechanical systems (MEMS). The system fails if the cumulative damage exceeds a time-dependent threshold. We develop a preventive maintenance policy to replace the system such that its lifetime is prudently utilized. Further, three variations on the healing pattern have been considered: (i) shocks heal for a fixed duration $\tau$; (ii) a fixed proportion of shocks are non-healable (that is, $\tau=0$); (iii) there are two types of shocks --- self healable shocks heal for a finite duration, and nonhealable shocks inflict a random system degradation. We implement a proposed preventive maintenance policy and compare the optimal replacement times in these new cases to that of the original case where all shocks heal indefinitely and thereby enable the system manager to take necessary decisions in generalized system set-ups.
\end{abstract}

% keywords can be removed

 \par \vspace{-1mm} 

\section{Introduction}
Industrial systems are often challenged by external impetus that affect their normal functioning. An impetus that inflicts a damaging effect is called a ``shock''. In the last few years, there have been extensive studies on different types of external shocks and their effects on a system. 
On the other hand, when an impetus produces a positive effect on the system by improving its current state, it is called a ``healing effect''. When the system heals by default, without requiring intervention, it is called ``self-healing''. Such natural and continuous self-healing is commonplace in many industrial, ecological, and biological systems and may continue either indefinitely or for a specific duration. 

Self-healing exists in software debugging systems, where bugs, malware invasion, license expiration, etc. are considered shocks, while automatic system cleansing is considered self-healing. Other industries such as polymer industries, and micro-electro-mechanical systems (MEMS) also observe self-healing phenomena. More details on MEMS are discussed in Subsection \ref{subsec}. Industries spend millions of dollars to monitor and maintain systems to prevent them from failing, especially when such a failure is catastrophic. These maintenance policies seek optimal rules to replace the system before risking its failure.

In recent decades, many shock models with \textit{healing} effects have been studied that also permit sporadic shocks of variable magnitudes and continuous internal degradation. For example, in Shen et al. (2018) \cite{shen2018system} shocks arrive according to a Poisson process with changing intensities. Depending on the degree of accumulated damage, the system performance can be divided into several states. In some states, self-healing reduces accumulated damage; however, self-healing can stop when the system reaches a specific state. There may be an internal degradation process also. For instance, in some systems or components like MEMS and servo motors as described in Wang et al. (2020) \cite{wang2020modeling}, natural degradation affects the consequences caused by shocks and vice versa. That paper allows a natural degradation state (NDS) function to classify shocks into safety, damage, and fatal zones according to their thresholds and derives a closed-form reliability function and failure time distribution function. 
Dong et al. (2020) \cite{dong2020reliability} introduced a ``damage recovery factor" to quantify self-healing and its effect on the reliability function and the mean failure time. They allow random shocks to accelerate internal degradation rate, and discuss a preventive replacement policy. Similarly, Kong et al. (2020) \cite{kong2020reliability} formulated a reliability model in multiple competing failure processes by considering the magnitudes of the shock and their duration simultaneously to study their impacts on the degradation processes, describing both recovery level and recovery time.
Ranjkesh et al. (2019)\cite{ranjkesh2019new}, evaluate system reliability using dependencies between inter-arrival times having phase-type distributions and shock magnitudes. Kang et al. (2022) \cite{kang2022reliability} provides a method for analyzing the reliability of systems that have self-healing mechanisms and are subject to cumulative shocks of two types --- fixed-interval shocks and random shocks ---following a certain inter-arrival time distribution, and show that self-healing mechanisms can significantly improve system reliability under both types of shocks. Hashemi et al. (2022) \cite{hashemi2022preventive} investigates maintenance strategies for repairable systems experiencing two distinct types of failure: internal aging-related failures and fatal external shocks following a nonhomogeneous Poisson process.

Systems exposed to shocks may also undergo \textit{degradation}, which has been studied extensively in the literature. For example, Dong et al. (2023) \cite{dong2023designing} provides a model for designing proactive replacement strategies for degraded systems subject to two types of external shocks: Proactive strategies involving replacing components before failure and reactive strategies involving waiting until a failure. Proactive strategies are generally more cost-effective than reactive strategies, especially in systems subject to external shocks. 
Cha and Finkelstein (2019) \cite{cha2019optimal} studies the operation of systems in a dynamic environment where the system hazard is represented as an additive hazard rate model, where shocks arrive according to homogeneous or nonhomogeneous Poisson Process.
Ye et al. (2023) \cite{ye2023generalized} presents a generalized dynamic stress-strength interference model for a self-healing protective structure that is subject to dynamic loads. The model is developed based on the $\delta$-failure criterion and takes into account the uncertainties in the stress and strength parameters and it is shown to accurately predict the failure probability of the self-healing protective structure under dynamic loading conditions. 
Chang et al. (2021)\cite{chang2021reliability} performs a reliability analysis method for systems that are subject to degradation over time by taking into account the change in the degradation rate and hard failure threshold as the system degrades. Moreover, Dong et al. (2021) \cite{dong2021reliability} also discusses internal degradation.

\par
In the literature, various kinds of policies are implemented to maintain the system. A survey of various maintenance policies for industrial systems is provided in Wang (2002) \cite{wang2002survey}, including a broad spectrum of replacement policies such as age-dependent, periodic, failure limit, and sequential preventive maintenance policies. These policies essentially use one of the following optimization criteria: maximize availability, minimize expected cost per unit time, minimize downtime, and minimize limiting failure rate. They also consider various repair policies such as perfect or imperfect repair, and various monitoring strategies such as at discrete time points or continuously. We will proceed to seek an optimal maintenance policy that minimizes the average cost per unit of time.
Preventive maintenance (PM) and corrective maintenance (CM) are the two classical types of maintenance policies undertaken to maximize profit or minimize loss due to failure. In Chien et al. (2012) \cite{chien2012optimal}, each period of operation inflicts a random amount of damage to the system and those damages accumulate to trigger a PM or a CM action.  The long-run expected cost rate is minimized to determine the optimal policy.
Qiu et al. (2020) \cite{qiu2020dynamic} develops a novel reliability model characterizing the self-healing effect on system reliability. They carry out an imperfect repair following each minor failure and replace the system based on its lifetime and the number of minor failures. The optimal replacement time is determined using a stochastic dynamic programming formulation that minimizes the expected total cost of system failure and imperfect repairs. In Dong et al. (2021) \cite{dong2021reliability}, external shocks and their damaging effects are considered for multi-component systems which are subjected to dependent and competing failure processes. Generalized shock models are presented under several shock categories. A block replacement policy is introduced and the Nelder–Mead downhill simplex method is employed to determine the optimal replacement interval based on the derived system reliability. Zhang and Yang (2020a) {\cite{zhang2020postponed} introduces a state-based postponement maintenance policy for asset management, considering uncertain environmental stresses and defect signals which address the challenges of optimizing the remaining useful lifetime by characterizing environmental damage and health state variations. It offers two levels of postponed maintenance windows based on inspection consequences.
In another work, Zhang and Yang (2020b) \cite{zhang2020state}, the authors developed a state-based maintenance policy with multifunctional maintenance windows, considering the impact of environmental disturbance on health variation and defect propagation. Here, three types of maintenance windows are scheduled, allowing flexible allocation of inspection and spare part resources.}

Stage-based stochastic models have also been considered in the literature for studying the reliability of a system. For example, Finkelstein and Gertsbakh (2016) \cite{ finkelstein2016preventive} considers two preventive maintenance strategies: one based on the system entering intermediate states between the initial UP state and the final absorbing DOWN state, and the other based on the occurrence of a certain number of shocks. Further, Finkelstein, Maxim and Eryilmaz (2021) \cite{finkelstein2021optimal} establishes conditions for the existence of a unique and finite preventive maintenance time based on the dynamic reliability characteristics of the system.
Applications of systems exposed to external shocks, internal degradation, and experiencing self-healing can be found in MEMS (See Dong et al. (2020) \cite{dong2020reliability} and  Dong et al. (2021) \cite{dong2021reliability}) and power transmission systems (see Kong et al. (2020) \cite{kong2020reliability}).

The current research builds upon Chatterjee and Sarkar (2022) \cite{chatterjee2022optimal} and Chatterjee and Sarkar (2021) \cite{Chatterjeecost}, where the system was exposed to randomly arriving external shocks of the same magnitude. Here, unlike the above papers, we let external shocks inflict damages of varying magnitudes. We also permit the system to start to heal instantaneously and continue to heal at a fixed rate while also continuously degrading due to aging. Under this more general model of variable damage, our objective is to determine an optimal replacement time that minimizes the cost per unit time. {The current work extends the above literature by not only considering either healing or degradation but both at the same time while employing condition-based replacement policies. We also do not restrict ourselves to any well-known arrival process for the various kinds of shocks but instead allow ourselves to consider any arbitrary distribution of inter-arrival times of those shocks. By building probabilistic models, we simulate various situations which are inspired by real-life systems such as the MEMS to guide system managers in their decision-making by weighing the risk of failure during the lifetime of the system.} 

The remainder of the paper is organized in the following way. Section~\ref{sec2} describes the evolution of the system as a continuous-time stochastic process that renews itself after a preventive or corrective replacement {and presents a practical industrial example where such a model may be useful}. Subsection~\ref{subsec1} explains the method of computing the expected cost per unit of time. Subsection~\ref{simulation_original} reports the optimal replacement times obtained in simulation studies. Furthermore, once we have established the optimal replacement policy, we also consider different variations of our assumed system in Section \ref{sec3}: First, the damaging shocks can heal only for a finite duration $\tau$ (Subsection \ref{tau_only}); second, a fixed proportion of shocks are non-healable (that is, $\tau= 0$ for these shocks) (Subsection \ref{nonhealable_only}); and third, there are two different types of shocks---healable for a finite duration $\tau$ and non-healable (Subsection \ref{types_of_shock}).  Details of the simulation studies for the three subcases are mentioned in Section \ref{sec3}. Section \ref{sec4} summarizes our research findings.

\section{Stochastic Evolution of Systems}\label{sec2}

The system described in this research is either a single- or a multi-unit system. External shocks arrive with random inter-arrival times inflicting damages of random magnitudes. Immediately after a shock arrives, the system begins to heal, which reduces the accumulated damage. We assume that exponential healing occurs continuously according to an exponential function and at a constant rate; therefore, cumulative damage decreases exponentially. %No additional stimulus is required for the system to heal: It heals naturally, continuously, and at a constant rate. 
The system fails when cumulative damage crosses a certain threshold, which decreases over time as a result of aging.

\textbf{The set-up and assumptions:}

\begin{enumerate}[label={(A\arabic*)}, align=left]
    \item Let $X_1,X_2,\dots,X_n$ denote the inter-arrival times of shocks which are {independently and identically distributed} (IID) with arbitrary CDF $F$.
    \item Let $Y_1, Y_2,\dots, Y_n$ denote the corresponding magnitudes of damage caused by the external shocks, which are IID with arbitrary CDF $G$.
    \item Damages from the shocks accumulate over time. 
\item The system self-heals from the damages at a constant rate. Hence, at any given time, either a shock arrives, causing the cumulative damage to shoot up, or the system continuously heals from the effects of all previous shocks, causing the cumulative damage to decrease continuously. {Instantaneous start of healing effect has been previously studied in the literature. In Liu et al. (2017) \cite{liu2017reliability}, healing started immediately and decreased linearly. Also assuming an instantaneous start of healing, Liu et al. (2018) \cite{liu2018reliability} extended the healing functions to include exponential healing for an indefinite duration, exponential healing for a limited duration, and a harmonic healing function.}

\item The system fails when the accumulated damage exceeds a certain boundary threshold, which decreases over time at a faster rate as the system ages, making it more vulnerable to failure.
{The justification behind such an assumption is that an aging system cannot tolerate as much damage as a younger system can. While many papers in the literature assume a constant threshold, some recent papers such as Liu et al. (2018) \cite{liu2018reliability} and Hao and Yang (2018) \cite{hao2018reliability}, consider variable thresholds based on the system condition at a given time. Because the damage-tolerance capacity of the system decreases over time, we assume the tolerance threshold is a non-increasing function.}
Here, for illustration, we assume that the boundary is a quadratically decreasing function. 
{We choose this function only for illustration purposes. The user may consider any other non-increasing function.}
Thus, the system fails in one of two ways:
\begin{enumerate}[label={(\roman*)}, align=left]
    \item a new shock arrives so that the cumulative damage exceeds the boundary;
    \item the accumulated damage, though decreasing, crosses the boundary while the system is healing because aging causes the boundary to come down faster.
\end{enumerate}
\end{enumerate}

\begin{figure}[ht]
    \centering
    \includegraphics[scale=0.50]{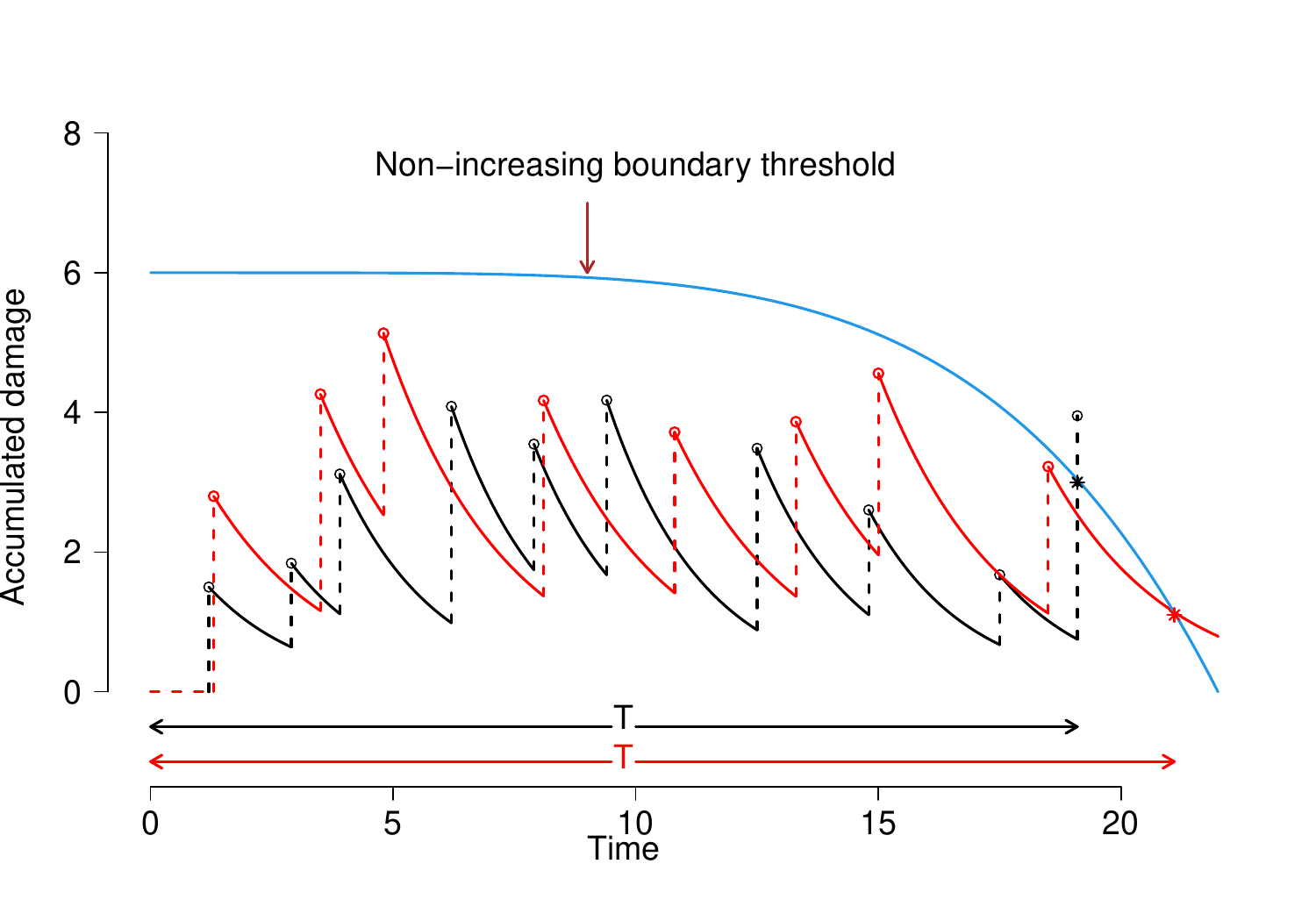}
    \caption{Depicting cumulative damage (black and red curves) as shocks arrive randomly. The blue curve represents the quadratically decreasing boundary threshold; dotted vertical segments denote the random amount of damage inflicted by each shock, and the continuous curves represent the exponential decay of cumulative damage due to constant healing. When the cumulative damage exceeds the boundary threshold, the system fails. }
    \label{sys}
\end{figure}

Figure \ref{sys}, which depicts the accumulated damage as a function of time, illustrates these two types of failure using two sample paths. The black sample path crosses the boundary threshold when a shock of sufficient magnitude arrives. The red sample path exceeds the boundary while the system is healing, but the boundary is reducing faster. 

\subsection{{\textbf{A motivating example}}}\label{subsec}

{The model discussed in this paper is a probabilistic approach to studying system reliability. Even though systems in real life may not exactly abide by the assumptions cited in this paper, inspired by some recent papers, we mention potential applications of such a system. Our motivation comes from the micro-electro-mechanical systems (MEMS) studied in}
Miller et al. (1997)\cite{miller1997routes}, Tanner and Dugger (2003) \cite{tanner2003wear}, Kim et al. (2007) \cite{kim2007nanotribology}, Liu et al. (2017) \cite{liu2017reliability}, Che et al. (2019) \cite{che2019reliability}, {for example}.

{MEMS utilize miniature devices that integrate mechanical and electrical components. MEMS are composed of mechanical microstructures, microsensors, microactuators, and microelectronics, all combined into a single silicone chip. Microsensors detect modifications in the system surroundings by measuring mechanical, thermal, magnetic, chemical, or electromagnetic variables. The integrated circuits are engineered to assimilate such information and respond appropriately by using the electrical and/or mechanical properties of silicone. The application of MEMS technology is widespread: It is used in inkjet printers, projection displays, data storage systems, optical and wireless telecommunications, smartphones, smartwatches, safety equipment such as airbag systems, etc. In biomedical fields, MEMS are widely used in ICU equipment, kidney dialysis units, scanning probe microscopes, DNA and RNA sequencers, etc. For such systems, maintenance personnel with proper domain knowledge will find the stochastic model and the computational techniques proposed in the upcoming sections useful for decision-making.}

{MEMS are susceptible to a wide range of random shocks that cause damage: mechanical shocks such as static friction, fracture, creep, and degradation of dielectrics; environmental shocks such as fluctuations in humidity; electrical shocks such as sudden voltage drops, etc. Sensors and testers can identify the arrival of shocks that affect the silicone layer of the chip. As soon as the system experiences a shock, lubricant films on the silicone surface initiate self-healing instantaneously and at a constant rate; see Kim et al. (2007) \cite{kim2007nanotribology}.}

{Suppose that from prior failure data, we have information on the initial threshold. Thereafter, as time progresses, chemical (corrosion, chemical reaction, etc.), mechanical (pressure, friction, etc.), and physical (wear debris on rubbing surfaces) changes reduce the fault tolerance threshold, captured by equation (\ref{thres}), for example.}

{Thus, MEMS are real-life examples of systems subject to shocks, degradation, and self-healing. The main contribution of this paper is to empower system managers to decide when to replace a system experiencing sporadic shocks arriving according to an arbitrary distribution, healing according to various patterns, and degrading due to aging.}

\subsection{Methodology}\label{subsec1}

Let us explain how to calculate the cumulative damage to the system at any given time: { Let $S_j = \sum_{i=1}^j X_i$ be the arrival time of the $j$-th shock. The damage amount of the first shock is $D_1 = Y_1$. This shock starts to heal instantaneously and continues to heal until the next shock arrives. Therefore, at time $t\, (>S_1)$, the accumulated damage is $D_1 e^{-\kappa(t-S_1)}$. As soon as the next shock arrives at time $t=S_2$ inflicting additional damage $Y_2$, the cumulative damage amount becomes $D_2=Y_2+ D_1 e^{-\kappa\, X_2}$.
For time $t\, (>S_2)$ the accumulated shock is $D_2 e^{-\kappa(t-S_2)}$, and so on.} Let $N(t)$ denote the number of shocks that have arrived by time $t$ (observed at increments of $\Delta$). Then the cumulative damage $D$, at time $t$ is computed as
\begin{equation}\label{damage}
 D(t) = \sum_{i=1}^{N(t)} Y_i e^{-\kappa(t-S_i)} 
 \end{equation}
where $Y_i$ is the damage inflicted by the $i$-th shock, and $\kappa$ is the fixed healing rate. 

For monitoring purposes, the system is observed at regular epochs at increments of $\Delta$ over a window of time $[0,\mathcal{T}]$. Borrowing the discretization approach in Chatterjee and Sarkar (2020) \cite{chatterjee2020computing}, we choose $\Delta$ sufficiently small so that for all practical purposes we observe the system almost continuously. 

{We consider a non-increasing boundary threshold $B(t)$. For illustration, we choose the following quadratically decreasing function }
\begin{equation}\label{thres}
    B(t) = a + b t - c t^2
\end{equation}
where $a, c \in \mathbb{R}^+$ and $b \in \mathbb{R}$. Note that, the above choice of the boundary threshold is illustrative only, one can consider any other non-increasing function of time. At each observation epoch, we measure the difference between the cumulative damage. For $k=1,2,\dots$, if the cumulative damage $D(t)$ does not cross the threshold $B(t)$ at the $(k-1)$-st observation epoch, but is found to have crossed it at the $k$-th inspection epoch, then the system has failed in between these two epochs, and we replace it at time $T=k\Delta$.  \par
\begin{figure}[ht]
    \centering
    \includegraphics[scale=0.60]{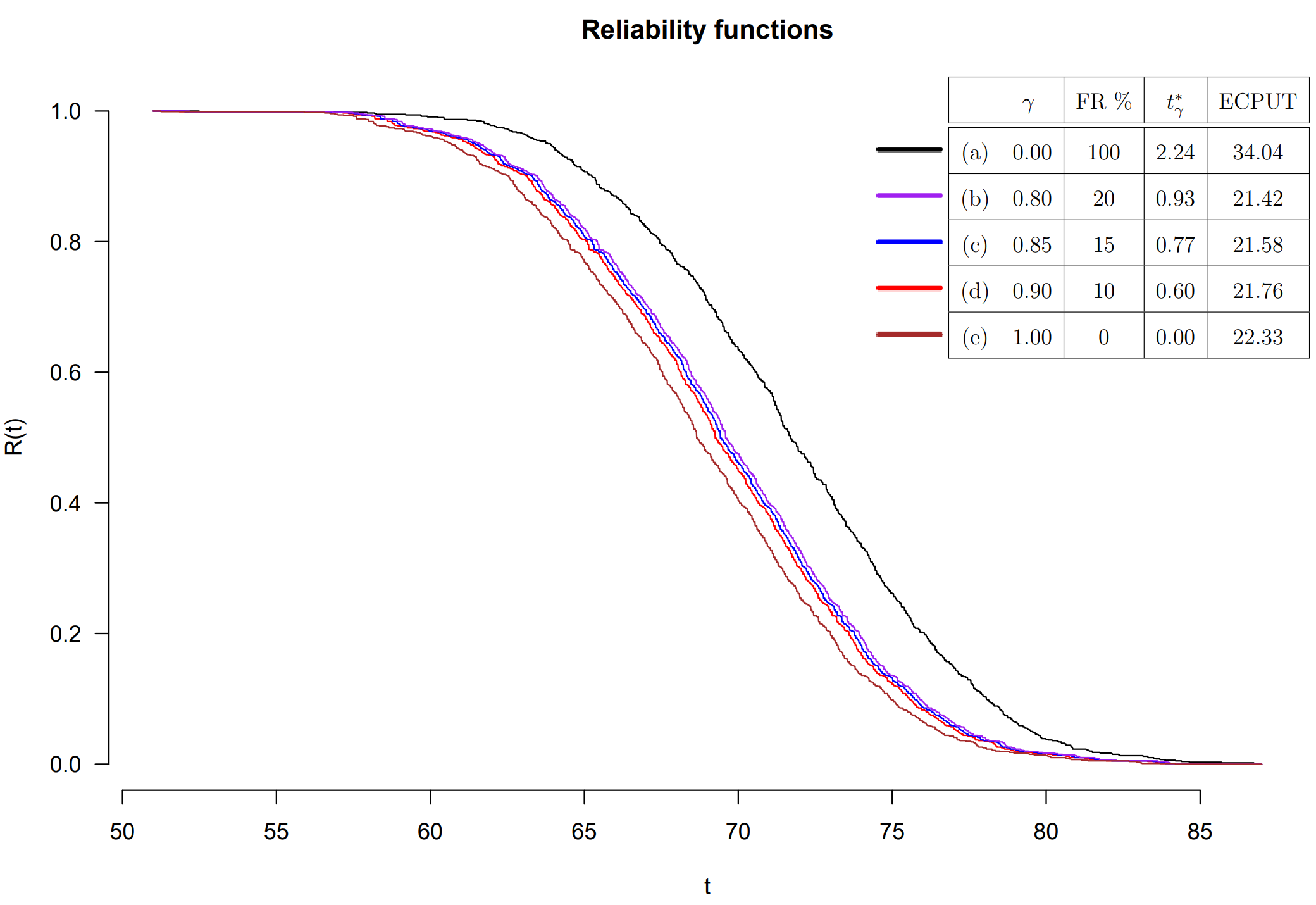}
    \caption{Reliability functions plotted against time for the displayed situations when $X \sim$ Weibull(2,$2/\pi$), $Y \sim$ Weibull(2,1/2), $B(t) = 500 - {t^2}/{50}$ and $\kappa = 0.02$. 
    }
    \label{rel}
\end{figure}

However, replacement upon failure is not desirable due to the high cost of replacing a failed system and the loss of revenue until a new system is installed. This has been explained in more detail in Figure \ref{rel}. Therefore, we must determine a preventive maintenance policy in which we replace the system before failure; however, we must not replace the system too early and forfeit its remaining lifetime. Therefore, we propose the following maintenance policy: Whenever the cumulative damage $D(t)$ enters within $d$ units of the threshold $B(t)$, where $d>0$ is yet to be determined, an alarm is activated and we replace the system after an additional time $t^*$ which depends on the tolerable risk probability (say, between 10\% and 20\%) that the system might fail before that epoch. Thus, the choice parameter $d$ is related to the additional duration $t^*$ after the alarm sets off when we replace the unfailed system.
For any choice of $d \in [D_1, D_2]$, at increments of $0.1$, we apply Algorithm~1 to compute lifetime, replacement time, residual lifetime, and the number of shocks.

{\bf Algorithm 1}
\begin{enumerate}[label={(S\arabic*)}, align=left]
        \item Generate $n$ shocks with inter-arrival time $X_1,X_2,\dots, X_n$ IID with CDF $F$ and magnitudes. $Y_1,Y_2,\dots,Y_n$ IID with CDF $G$, where $n$ is sufficiently large; say, $n\approx 2\,  \mathcal{T}/E[X]$.
        \item Calculate cumulative damage $D(t)$ at each epoch $t=j\Delta$ (for $j = 1,2,\cdots$) using equation (\ref{damage}). Suppose that $T=k\Delta$ is the first time $D(t)$ exceeds the boundary. Then we must replace the system at epoch $T$, called \textit{failure time}. 
        \item We record the number $N$ of shocks that the system endures until failure.
        \item Let $T' =l\Delta$, for some $l<k$, be the first time that cumulative damage $D(t)$ is within $d$ units of the boundary. Had we replaced the system at $T'$, the lifetime of the system lost due to premature replacement would have been $T-T'$, and would be called \textit{residual lifetime}. We call $T'$ the \textit{premature replacement time}.
        \item Repeat steps (S1) to (S4) $10^4$ times. For each repetition, obtain $T$, $T'$, $N$, and compute $r=T-T'$.    
\end{enumerate}

Note that our objective is to utilize the system to its fullest. Therefore, we should not replace the system too soon. How long could we allow the system to function before replacing it so that the chance of a system failure within this additional duration would be 10\%, 15\%, or 20\% (equivalently, survival probabilities would be 90\%, 85\% or 80\%), respectively? Note that we are only allowing up to a 20\% risk of failure --- the reason behind that will be presented later. Let us denote these survival percentiles after the alarm sets off by $t_{.90}^*(d)$, $t_{.85}^*(d)$, and $t_{.80}^*(d)$, respectively (collectively denoted by $t_{\gamma}^*(d)$ for $\gamma=0.90, 0.85, 0.80$). We compute these survival percentiles using the $1000$ values of the residual lifetime $r=T-T'$.
This we do for every choice of $d \in [D_1, D_2]$, at increments of $0.1$.
Our main objective is to find an optimal $d$ such that the expected cost per unit of time is minimized when we are willing to risk a small (10\%, 15\% or 20\%) chance of system failure within the next $t^*_{\gamma}(d)$ units of time after the alarm sets off. 

To compute the expected cost per unit time, let $c_0$ be the initial cost of installing the system, $c_I$ be the per unit time cost of inspection and maintenance (although the inspection cost is incurred at increments of $\Delta$, we redistribute the cost over the entire interval), $c_{op}$ be the per unit time cost of operating the system, $c_{rev}$ be the per unit time revenue earned by the system while operating (it is a negative cost), and $c_f$ be the additional cost of failure replacement. 
{Here, we do not incur any cost to repair. We install a brand new unit and remove a failed/significantly damaged system and replace it with another new system. We only observe the system and do not intervene unless a system fails or is significantly damaged; hence, no repair is made. }
We define as cycle time the duration from the time a system is installed for operation until it is replaced at epoch $\min\{T, T' + t_{\gamma}^*(d)\}$. Then, the expected cost $(EC)$ within a cycle time is:
\begin{equation}\label{ec1}
{EC}[t_{\gamma}^*(d)] = c_0 + (c_I + c_{op} - c_{rev}) \times E[\min\{T,T' + t_{\gamma}^*(d)\}] + c_f I_F
\end{equation}
where $I_F=1$ if the system experiences failure and $I_F=0$
if the system is replaced before failure. 
We wish to minimize ${EC[t_{\gamma}^*(d)]}/{E[\min\{T, T' + t_{\gamma}^*(d)\}]}$ to get expected cost per unit time (ECPUT) with respect to $d$.

\subsection{Simulations}\label{simulation_original}
To demonstrate our proposed policy for finding the optimal $d$ and associated expected $t_{\gamma}^*(d)$, we choose  $\mathcal{T} = 100, \Delta=0.05$, $\kappa = 0.01$ or $0.02$, $E(X)=1$ and $E(Y)=10$, respectively. We take  $n=200$ so that in every iteration, the cumulative damage (almost) surely crosses the boundary (\ref{thres}) with $a = 500$, $b=0$ and $c = 0, 1/60, 1/50, 1/40, 1/30$. We calculate the cumulative damage at time $t$ using equation (\ref{damage}). We make 1000 such iterations to obtain our results. Let the various costs be $c_0 = 5000, c_I = 50, c_{op} = 100, c_{rev}=200, c_f = 1000$. We search for optimal $d \in [D_1=8, D_2=16]$. \par

{In Figure \ref{rel}, we present the reliability diagram for various cases of replacing the system. When the system is removed after a failure, we call it a 100\% failure risk (FR) situation. If the system is replaced as soon as the cumulative damage reaches within $d$ units of the boundary for the first time but has not failed, we call it a 0\% risk situation. When cumulative damage reaches within $d$ units of the boundary and we allow the system to function for an additional $t_{\gamma}^*$ time, where $(\gamma = 0.90,0.85,0.80)$, we call them 10\%, 15\% and 20\% risk situations, respectively. After calculating the ECPUT for these situations, we plot the reliability functions for each of them. We see that the utilization time of the system is the highest if we replace it at failure, but it has a high cost. On the other hand, the reliability is the lowest for the 0\% risk situation; but the cost incurred is much lower. We see that $t_\gamma^*$ for $\gamma=0.80$ is almost half of the mean residual lifetime. When we are willing to take a 20\% risk of failure, the ECPUT is the smallest. However, it is only 1.1\% lower than the corresponding ECPUT at 10\% risk. Therefore, increasing the risk further is not worthwhile. A low-risk situation (10\% or 15\%) is a good trade-off between reliability and cost. Thus, reliability should not be the sole criterion for deciding when to replace the system; we must also minimize the cost. Due to similar results throughout, we base our analysis on minimizing the cost per unit of time in the following sections.}\par

\begin{figure}[ht]
    \centering
    \includegraphics[scale=0.50]{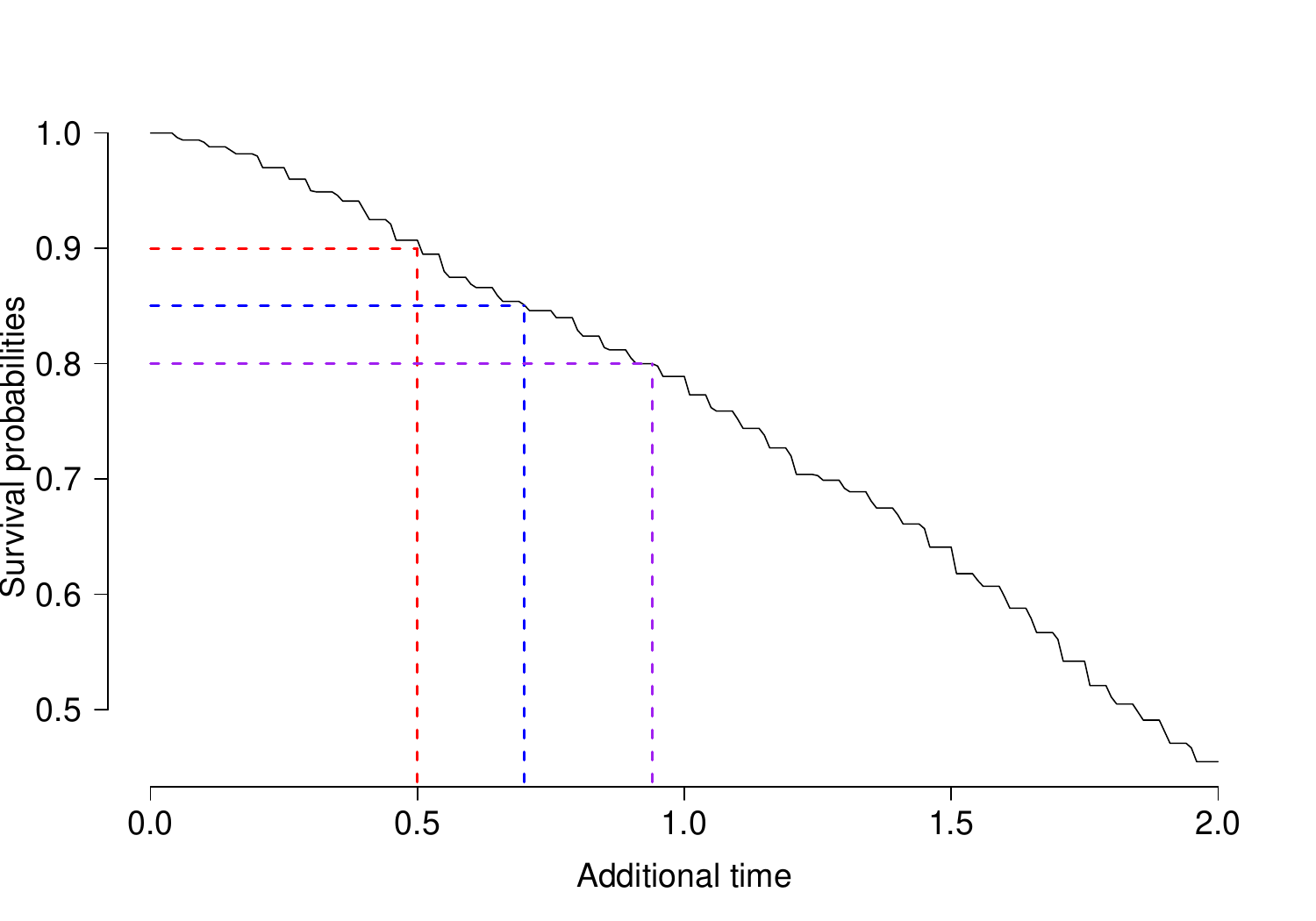}
    \caption{ When $B(t) = 500 - {t^2}/{50}$, $\kappa = 0.02$ and $d=12$, for $\gamma = 0.80, 0.85$ and $0.90$,  the additional time $t^*_{\gamma}$ that the system should be allowed to operate after the alarm sets off are $0.94, 0.70$ and $0.50$ units of time respectively.}
    \label{surv}
\end{figure}

For inter-arrival time between shocks, we choose $X \sim$ Weibull (shape = 2, scale = $2/\sqrt{\pi}$) so that $E(X)=1$. For magnitude of shocks, we choose $Y \sim$ Weibull (shape = 10, scale = $50/\Gamma(1/5)$) so that $E(Y) = 10$.
For each choice of $d\in  [8,16]$ at increments $0.1$, values of $t^*_{\gamma}$ are obtained for $\gamma = 0.90, 0.85, 0.80$. In Figure \ref{surv}, as an illustration, we display the survival plot for a particular choice $d=12$ showing that  $t_{.90}^*=0.50, t_{.85}^*=0.70, t_{.80}^*=0.94$. 

\begin{figure}[ht]
    \centering
    %\subfloat[\centering ]
    \includegraphics[scale=2.00]{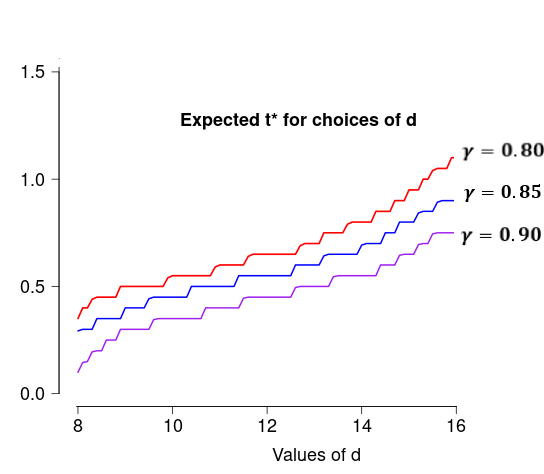} 
    %\qquad
    %\subfloat[\centering ]
    \caption{The percentiles $t^*_{80} \geq t^*_{85} \geq t^*_{90}$ are increasing functions of $d$. }
    \label{d_vs_tstar}
\end{figure}

%\newpage
\begin{figure}[ht]
    \centering
    %\subfloat[\centering ]
    %\qquad
    %\subfloat[\centering ]
    \includegraphics[scale=0.50]{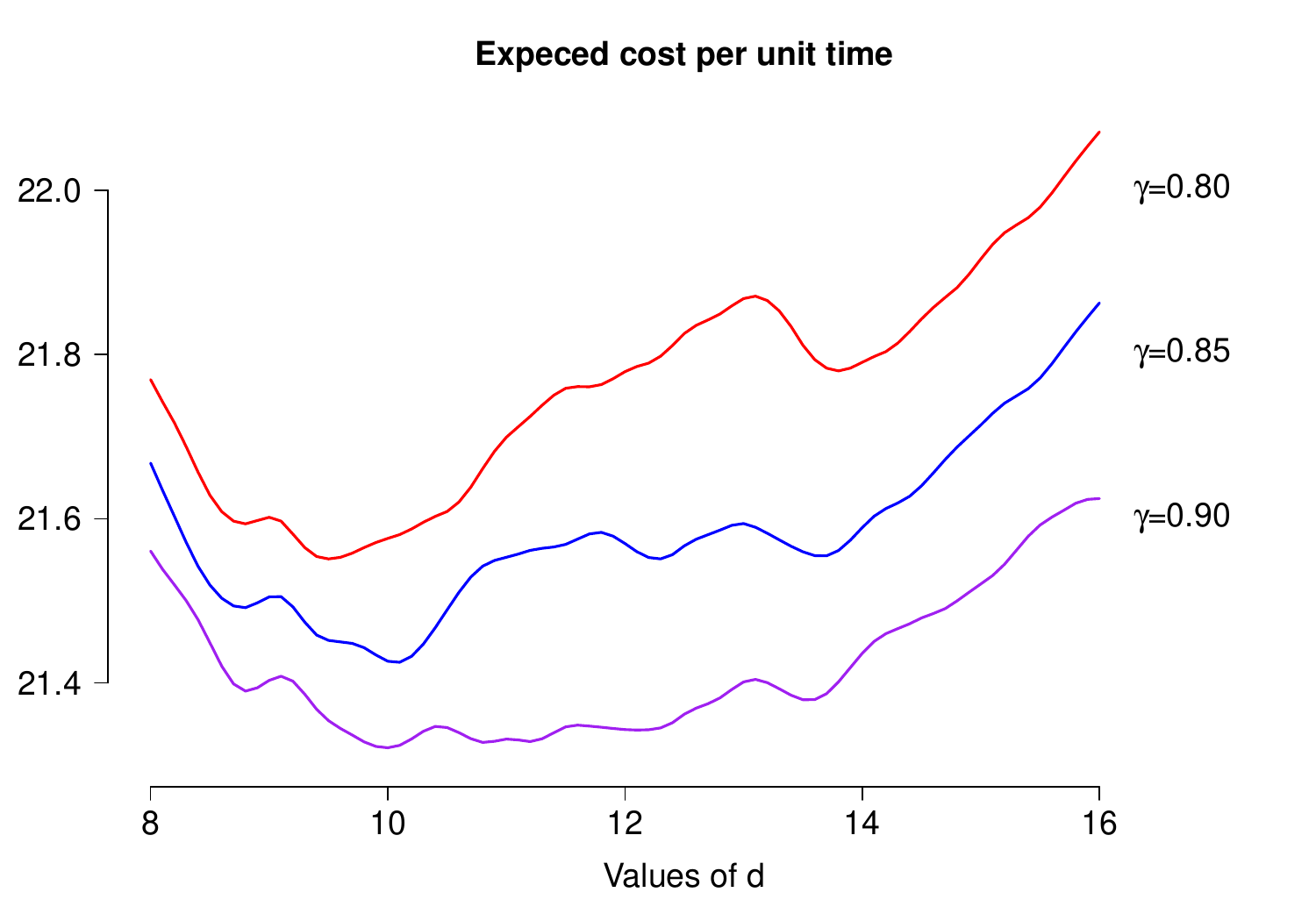}%
    \caption{The expected cost per unit time is minimized at $d = 9.4$ for $\gamma = 0.80$, at $d = 10.1$ for $\gamma = 0.85$, and at $d=10.0$ for $\gamma = 0.90$. If there are multiple minima, choose the smallest one.}
    \label{d_policy_original}
\end{figure}

Next, using equation (\ref{ec1}), the expected cost per unit time is calculated for every choice of $d$ and the associated $t^*_{\gamma}$, for  $\gamma=0.90, 0.85, 0.80$.
Figure \ref{d_vs_tstar} demonstrates that the larger the $d$ (that is, the farther the accumulated damage from the boundary), the larger the $t^*_{\gamma}$, for each $\gamma$. Figure \ref{d_policy_original} shows the expected cost per unit of time as a function of $d$ 
when $X \sim$ Weibull (shape = 2, scale = $2/\sqrt{\pi}$), $Y \sim$ Weibull (shape = 2, scale =1/2) and $B(t) = 500 - {t^2}/{50}$. If ECPUT has multiple minima, we choose the smallest one, since we wish to utilize the system as much as possible without compromising the cost per unit time. 

\begin{table}[ht]
\centering
\caption {For $X \sim$ Weibull(2,$2/\pi$), $Y \sim$ Weibull(2,1/2), and for various choices of $\kappa$ and $B(t)$, the optimal $d$ and [the associated $t^*_{\gamma}$] are displayed for $\gamma=0.80, 0.85, 0.90$. \\}
\begin{tabularx}{0.8\textwidth} {
  | >{\raggedright\arraybackslash}X 
  | >{\centering\arraybackslash}X 
  | >{\centering\arraybackslash}X 
  | >{\centering\arraybackslash}X | }
  \hline
  \multicolumn{4}{|c|}{$B(t) = 500$} \\
 \hline
 $\kappa$ & $\gamma = 0.80$ & $\gamma = 0.85$ & $\gamma = 0.90$\\
 \hline
 $0.01$   &  $~9.1 ~[0.625]$   &$~9.1~[0.803]$&    $~9.6~[0.963] $\\
 \hline
 $0.02$   & $~8.2~[1.105]$   &$~8.2 ~[1.137]$&   $~8.8 ~[1.423]$\\
 \hline
 \hline
 \multicolumn{4}{|c|}{$B(t) = 500- t^2/60$} \\
 \hline
$\kappa$ & $\gamma = 0.80$ & $\gamma = 0.85$ & $\gamma = 0.90$\\
 \hline
 $0.01$   & $~9.6 ~[0.448]$   &$~9.6 ~[0.583]$&   $~9.6 ~[0.709]$\\
 \hline
 $0.02$   & $~8.6 ~[0.610]$   &$~8.6 ~[0.782]$&   $~8.7 ~[0.953] $\\
 \hline
 \hline
 \multicolumn{4}{|c|}{$B(t) = 500- t^2/50$} \\
 \hline
 $\kappa$ & $\gamma = 0.80$ & $\gamma = 0.85$ & $\gamma = 0.90$\\
 \hline
 $0.01$   & $10.3 ~[0.456]$   &$10.4 ~[0.575]$&   $10.4 ~[0.679]$\\
 \hline
 $0.02$   & $~9.4 ~[0.601]$   &$10.1 ~[0.774]$&   $10.0 ~[0.934]$\\
 \hline
 \hline
 \multicolumn{4}{|c|}{$B(t) = 500- t^2/40$} \\
 \hline
$\kappa$ & $\gamma = 0.80$ & $\gamma = 0.85$ & $\gamma = 0.90$\\
 \hline
 $0.01$   & $10.6 ~[0.401]$   &$10.6 ~[0.534]$&   $10.6 ~[0.638]$\\
 \hline
 $0.02$   & ~$9.5 ~[0.530]$   &$10.2 ~[0.698]$&   $10.2 ~[0.843]$\\
 \hline
 \hline
 \multicolumn{4}{|c|}{$B(t) = 500- t^2/30$} \\
 \hline
 $\kappa$ & $\gamma = 0.80$ & $\gamma = 0.85$ & $\gamma = 0.90$\\
 \hline
 $0.01$   & $10.8~[0.376]$   &$10.8~[0.495]$&   $10.8~[0.606]$\\
 \hline
 $0.02$   & $~9.7 ~[0.526]$   &$10.7~[0.681]$&   $10.7~[0.813]$\\
 \hline
 \end{tabularx}
\label{table_optd}
\end{table}

Table \ref{table_optd} displays the simulation results showing the optimal $d$ for different choices of $\kappa$ and boundary thresholds with different quadratic coefficients, but keeping the inter-arrival distribution Weibull and the magnitude of shocks Weibull. 

Here are some lessons learned from Table \ref{table_optd}:

\begin{enumerate}
    \item 
If the healing rate $\kappa$ increases,  the system heals faster so that it takes longer for the accumulated damage to come within $d$ units of the boundary, thereby increasing the replacement time. By the same logic, a higher $\kappa$ also causes the optimal $d$ to be smaller. 

\item
Also, for a fixed $\kappa$ and for a particular choice of boundary threshold, if survival probabilities are chosen to be smaller, then $t^*$ increases or at least remains the same (because, by definition, $t^*_{80} \geq t^*_{85} \geq t^*_{90}$), and the optimal $d$ also increases, because of monotonic relation with $t^*$ exhibited in Figure~\ref{d_vs_tstar}. Conversely, if we demand a higher survival rate, then the optimal $d$ decreases.

\item
As the boundary threshold decreases at a faster rate, the optimal $d$ becomes larger and the corresponding $t^*$ is smaller for each of the $90\%$, $85\%$, and $80\%$ survival rates. This is because when the boundary decreases faster, we should let the alarm go off earlier to avoid potential failure. 

\end{enumerate}

For boundary $B(t) = 500 - t^2/50$, healing rate $\kappa=0.02$, and $Y$ following Weibull distribution,
Table~\ref{optd_dist_alt} shows the optimal $d$ and the associated $t^*_{\gamma}$ for various distributions of inter-arrival time of shocks. 

\begin{table}[ht]
\centering
\caption {For $Y \sim$ Weibull(2,1/2), $\kappa = 0.02$, and $B(t) = 500 - t^2/50$, and various inter-arrival time distributions, the optimal $d$ and [the associated $t^*_{\gamma}$] are displayed for $\gamma=0.90, 0.85, 0.80.$\\}
\begin{tabularx}{0.8\textwidth} { 
  | >{\centering\arraybackslash}X 
  | >{\centering\arraybackslash}X 
 | >{\centering\arraybackslash}X | }
 \hline
 \multicolumn{3}{|c|}{$X \sim gamma~(shape = 3, scale = 1/3) $} \\
 \hline
 $\gamma = 0.90$ & $\gamma = 0.85$ & $\gamma = 0.80$\\
 \hline
 $8.7 ~[0.542]$   &$8.7 ~[0.706]$&   $8.4 ~[0.865]$\\
 \hline
 \hline
 \multicolumn{3}{|c|}{$X \sim Weibull~(shape = 2, scale = 2/\sqrt{\pi})$} \\
 \hline
 $\gamma = 0.90$ & $\gamma = 0.85$ & $\gamma = 0.80$\\
 \hline
 $9.3 ~[0.601]$   &$10.1 ~[0.774]$&   $10.1 ~[0.934]$\\
 
 \hline
 \hline
 \multicolumn{3}{|c|}{$X \sim \textit{inverse-Gaussian} ~(mean = 1)$} \\
 \hline
 $\gamma = 0.90$ & $\gamma = 0.85$ & $\gamma = 0.80$\\
 \hline
 $9.9 ~[0.334]$   &$9.9 ~[0.432]$&   $10.5 ~[0.529]$\\
 
 \hline
 \hline
 \multicolumn{3}{|c|}{$X \sim exponential~(rate = 1)$} \\
 \hline
 $\gamma = 0.90$ & $\gamma = 0.85$ & $\gamma = 0.80$\\
 \hline
 $11.4 ~[0.143]$   &$10.9 ~[0.229]$&   $10.7 ~[0.315]$\\
 \hline
 \end{tabularx}
 
\label{optd_dist_alt}
\end{table}

When $X$ has Weibull or gamma distribution, the additional time  $t^*(\gamma)$ after the alarm sets off, is comparable, and the optimal $d$ is robust around $10$. However, when $X$ is inverse-Gaussian, the $t^*_{\gamma}$ values are much lower, because inverse-Gaussian distribution has a heavier right tail than Weibull and gamma. Likewise, because the exponential distribution has an even thicker right tail, the corresponding $t^*_{\gamma}$'s are even smaller. Among all inter-arrival time distributions considered here, exponential is the most heavy-tailed; hence its survival function is the highest, and the optimal $d$ is the largest.
In the next section, we discuss some variations in the healing pattern. 

\section{Variations in the healing effect and the shock types}\label{sec3}

In this section, we discuss some variations on the stochastic modeling of the system evolution described in Section \ref{sec2}.  

\subsection{Case~1: Healing stops after a finite duration}\label{tau_only}

\begin{figure}[ht]
    \centering
    \includegraphics[scale=0.50]{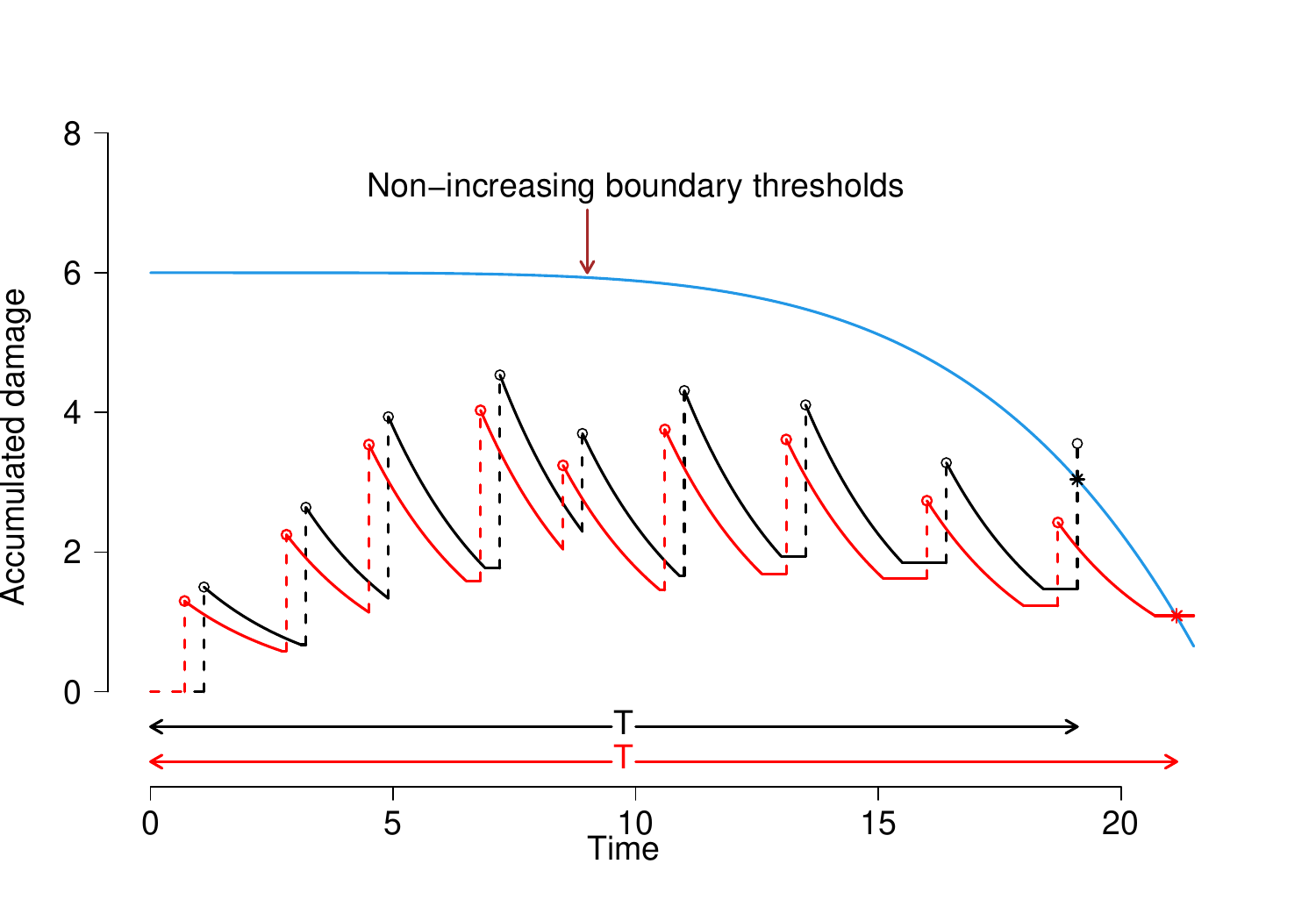}
    \caption{Depicting cumulative damage (black and red) as shocks arrive randomly. The blue curve represents the quadratically decreasing boundary threshold; dotted vertical segments denote a random amount of damage inflicted by each shock, and the continuous curves represent the exponential decay of cumulative damage up to a finite duration $\tau=1$ due to constant healing. When the cumulative damage exceeds the boundary threshold, the system fails.}
    \label{system_subcase1}
\end{figure}

Unlike in the previous section where healing continues indefinitely so that the damage eventually heals $100\%$, in this subsection, healing continues only up to a finite duration $\tau$, and thereafter stops, so that only a certain percentage of the inflicted damage heals. Conversely, if we specify what proportion of the inflicted damage will heal, we can find the corresponding $\tau$. Thus, the shocks are not totally healable, and a residual damage is left behind. Figure \ref{system_subcase1} illustrates the accumulated damage until system failure. 

For illustration, we make the following choices: For exponential healing with rate $\kappa=0.01$, we choose $\tau=50$ to attain a 40\% healing of the inflicted damage and $\tau=25$ for a 22\% healing. 
When the healing rate increases to $\kappa=0.02$, a choice of $\tau=25$ attains a $40\%$ healing, and $\tau=50$ attains a $64\%$ healing. {Using the notation $A \wedge B$ to denote the minimum of $A$ and $B$, the cumulative damage $D(t)$ at time $t$ is calculated as}

\begin{equation}\label{damage2}
  D(t) = 
 \sum_{i=1}^{N(t)} Y_i e^{-\kappa[(t-S_i)\wedge\tau]} 
\end{equation}

We see that the equation (\ref{damage2}) matches equation (\ref{damage}) when we let $\tau =\infty$. 
As in the previous section, we calculate the lifetimes, replacement times, and associated $t_\gamma^*$' s for survival probabilities $\gamma = 0.90, 0.85, 0.80$ after $10^4$ repetitions of the stochastic process. 
We implement the same preventive maintenance policy as in Subsection \ref{simulation_original} to document in Table~\ref{table_optd_subcase1_50} and Table~\ref{table_optd_subcase1_25} the optimal $d$ and the associated $t_\gamma^*$ under the modified healing rule for $\tau=50$ and $25$ respectively.

\begin{figure}[ht]
    \centering
    {{\includegraphics[width=0.70\textwidth]{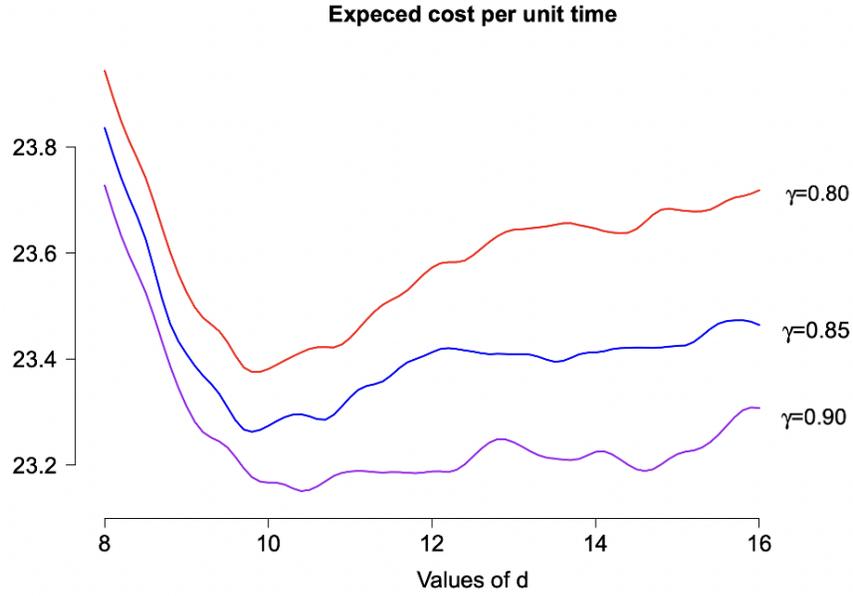} }}
\caption{With $\kappa=0.02$ and $\tau=50$, the expected cost per unit time is minimized at $d = 9.6$ for $\gamma = 0.80$ and $\gamma = 0.85$ and at $d=10.6$ for $\gamma = 0.90$.}
    \label{d_policies_tau_a}
\end{figure}

\begin{figure}[ht]
    \centering
    {{\includegraphics[width=0.70\textwidth]{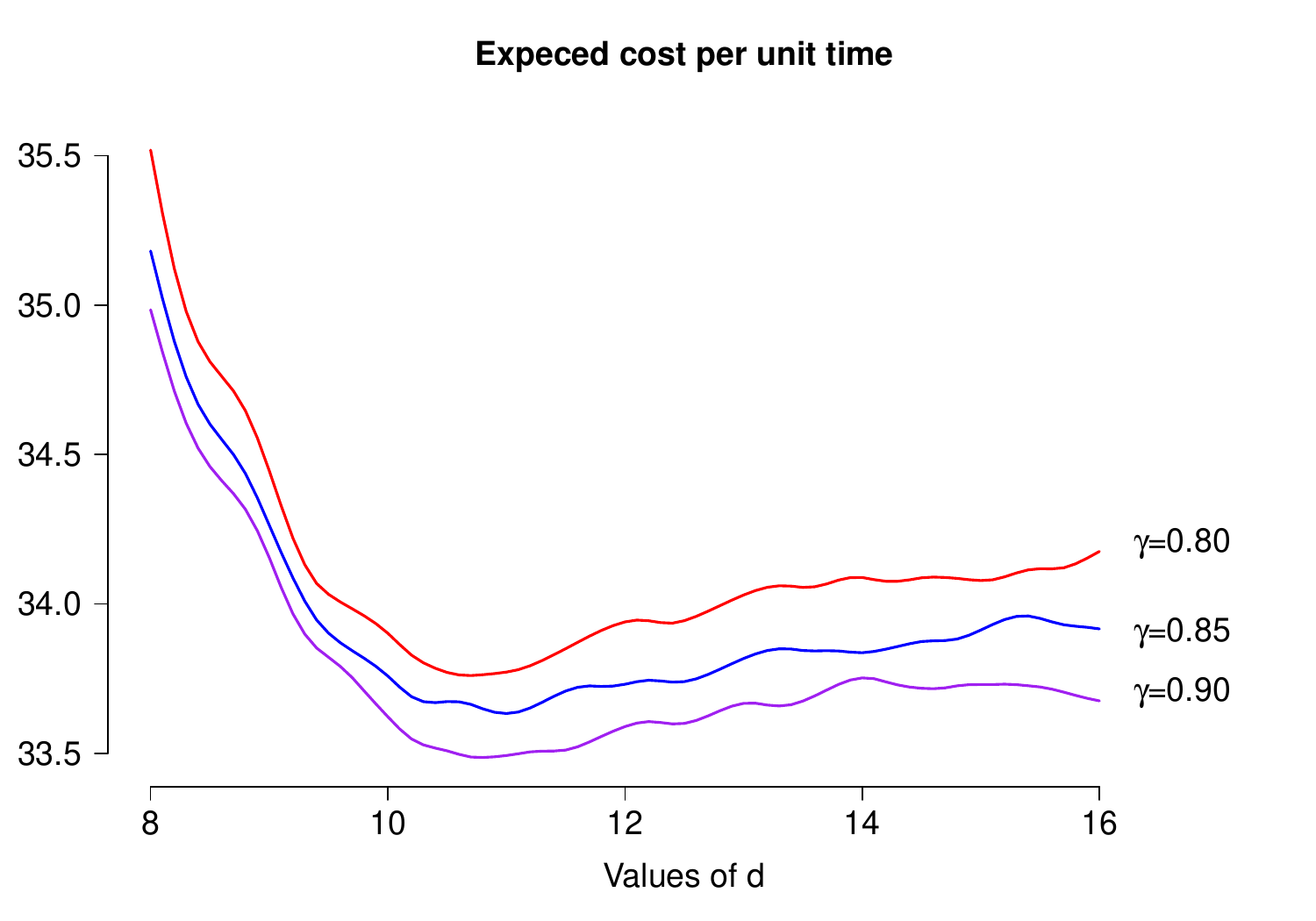} }}%
    \caption{ With $\kappa=0.02$ and $\tau=25$, the expected cost per unit time is minimized at $d = 10.6$ for $\gamma = 0.80$, $d = 10.4$ for $\gamma = 0.85$, and $d = 10.7$ for $\gamma = 0.80$. If there are multiple minima, choose the smallest one.}
    \label{d_policies_tau_b}
\end{figure}

\begin{table}[ht]
\centering
 \caption {For $\tau=50$,  $X \sim$ Weibull(2,$2/\pi$), $Y \sim$ Weibull(2,1/2), and for various choices of $\kappa$ and $B(t)$, the optimal $d$ and [the associated $t^*_{\gamma}$] are displayed for $\gamma=0.80, 0.85, 0.90$. \\}
\begin{tabularx}{0.8\textwidth} {
  | >{\raggedright\arraybackslash}X 
  | >{\centering\arraybackslash}X 
  | >{\centering\arraybackslash}X 
  | >{\centering\arraybackslash}X | }
  \hline
   \multicolumn{4}{|c|}{$B(t) = 500- t^2/60$} \\
 \hline
 $\kappa$ & $\gamma = 0.80$ & $\gamma = 0.85$ & $\gamma = 0.90$\\
 \hline
 $0.01$   & $~9.9 ~[0.447]$   &$~9.9 ~[0.575]$&   $~11.1 ~[0.690]$\\
 \hline
 $0.02$   & $~9.3 ~[0.578]$   &$~9.8 ~[0.742]$&   $~9.8 ~[0.880] $\\
 \hline
  \hline
  \multicolumn{4}{|c|}{$B(t) = 500- t^2/50$} \\
  \hline
 $\kappa$ & $\gamma = 0.80$ & $\gamma = 0.85$ & $\gamma = 0.90$\\
  \hline
  $0.01$   & $10.5 ~[0.435]$   &$11 ~[0.564]$&   $11.3 ~[0.683]$\\
 \hline
 $0.02$   & $~9.6 ~[0.552]$   &$9.6 ~[0.714]$&   $10.6 ~[0.863]$\\
  \hline
  \hline
 \multicolumn{4}{|c|}{$B(t) = 500- t^2/40$} \\
 \hline
$\kappa$ & $\gamma = 0.80$ & $\gamma = 0.85$ & $\gamma = 0.90$\\
  \hline
  $0.01$   & $10.6 ~[0.392]$   &$10.6 ~[0.527]$&   $11.4 ~[0.634]$\\
 \hline
  $0.02$   & ~$9.7 ~[0.505]$   &$10.1 ~[0.642]$&   $10.2 ~[0.774]$\\
 \hline
 %\hline
 \end{tabularx}
 \label{table_optd_subcase1_50}
\end{table}

%\newpage
\begin{table}[ht]
\centering
\caption {For $\tau=25$, $X \sim$ Weibull(2,$2/\pi$), $Y \sim$ Weibull(2,1/2), and for various choices of $\kappa$ and $B(t)$, the optimal $d$ and [the associated $t^*_{\gamma}$] are displayed for $\gamma=0.80, 0.85, 0.90$. \\} 
\begin{tabularx}{0.8\textwidth} {
  | >{\raggedright\arraybackslash}X 
  | >{\centering\arraybackslash}X 
  | >{\centering\arraybackslash}X 
  | >{\centering\arraybackslash}X | }
  \hline
   \multicolumn{4}{|c|}{$B(t) = 500- t^2/60$} \\
 \hline
 $\kappa$ & $\gamma = 0.80$ & $\gamma = 0.85$ & $\gamma = 0.90$\\
 \hline
 $0.01$   & $10.8 ~[0.338]$   &$11.3 ~[0.449]$&   $11.3 ~[0.568]$\\
 \hline
 $0.02$   & $10.3 ~[0.426]$   &$10.3 ~[0.562]$&   $10.3 ~[0.682] $\\
 \hline
  \hline
  \multicolumn{4}{|c|}{$B(t) = 500- t^2/50$} \\
  \hline
 $\kappa$ & $\gamma = 0.80$ & $\gamma = 0.85$ & $\gamma = 0.90$\\
  \hline
  $0.01$   & $11.3 ~[0.349]$   &$11.3 ~[0.456]$&   $11.3 ~[0.561]$\\
 \hline
 $0.02$   & $10.6 ~[0.430]$   &$10.4 ~[0.560]$&   $10.7~[0.669]$\\
  \hline
  \hline
 \multicolumn{4}{|c|}{$B(t) = 500- t^2/40$} \\
 \hline
 $\kappa$ & $\gamma = 0.80$ & $\gamma = 0.85$ & $\gamma = 0.90$\\
  \hline
  $0.01$   & $11.3 ~[0.339]$   &$11.5 ~[0.442]$&   $11.7 ~[0.541]$\\
 \hline
  $0.02$   & ~$10.8 ~[0.394]$   &$11.1 ~[0.517]$&   $11.1 ~[0.626]$\\
 \hline
 %\hline
 \end{tabularx}
\label{table_optd_subcase1_25}
\end{table}

Note that the overall optimal replacement time $t_\gamma^*$ when $\tau$ is a finite number is lower than that in Subsection~\ref{simulation_original} where $\tau=\infty$. This is anticipated because when the shocks do not heal indefinitely, their residual damages bring the collective damage closer to the boundary threshold much earlier. In general, we also see that the $t_\gamma^*$'s are significantly lower and the optimal $d$'s are larger in Tables \ref{table_optd_subcase1_50} and \ref{table_optd_subcase1_25} as compared to Table \ref{table_optd}, implying that the alarm goes off when the distance from the boundary is larger and we wait a shorter duration after the alarm goes off to replace the system. Further, comparing Tables \ref{table_optd_subcase1_50} and \ref{table_optd_subcase1_25}, we see that for the latter one, the optimal $d$'s are larger and the expected $t_\gamma^*$'s are lower because when $\tau=25$, the healing continues for a shorter duration than when $\tau=50$.

\subsection{Case~2: Some shocks are not healable}\label{nonhealable_only}

\begin{figure}[ht]
    \centering
    \includegraphics[scale=0.50]{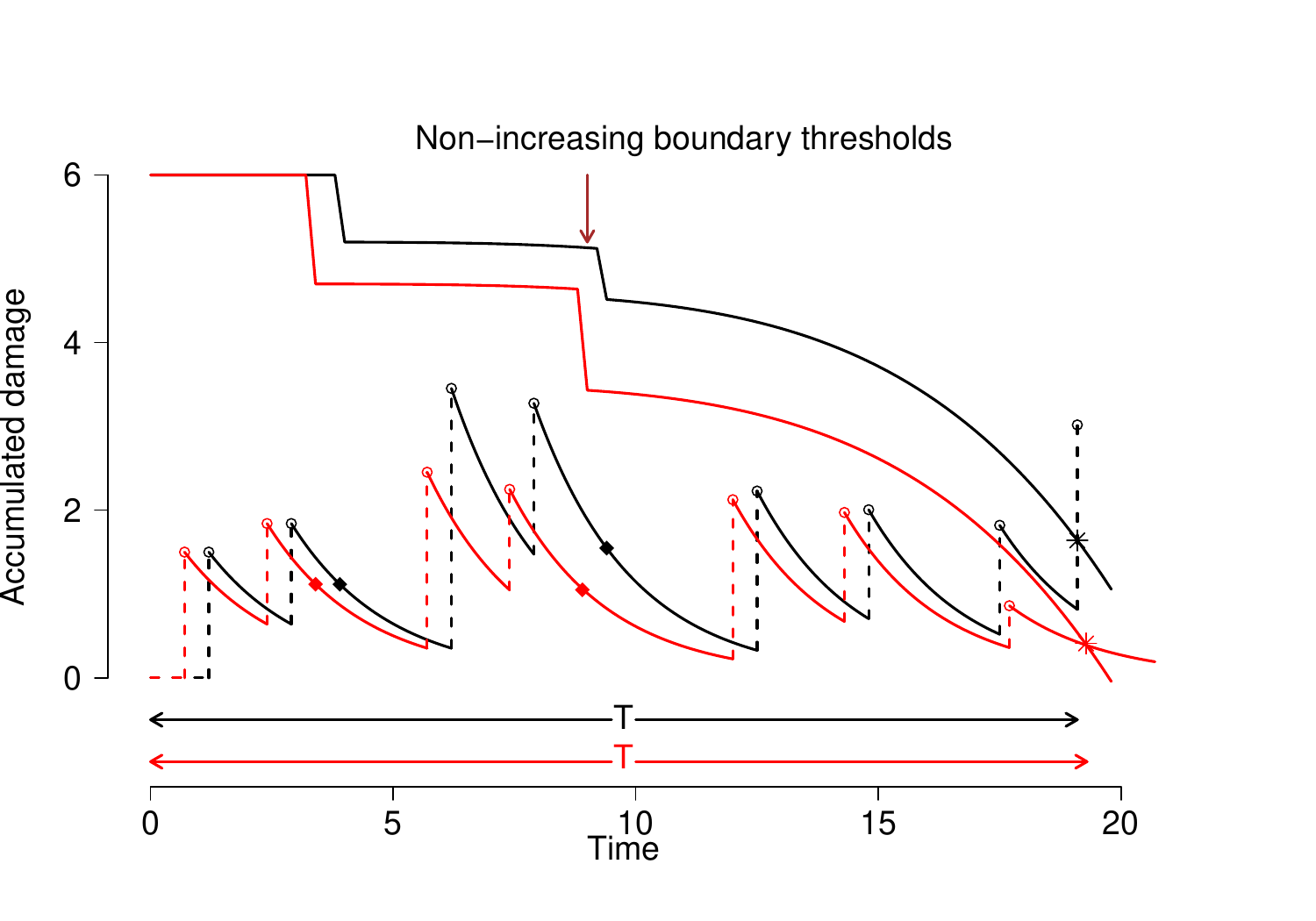}
    \caption{Depicting cumulative damage and the corresponding boundary curves (black and red curves) as shocks arrive randomly.  
    The boundaries drop due to the arrival of nonhealable shocks denoted by diamond-shaped dots on the stochastic paths.}
    \label{system_subcase2}
\end{figure}

In this subsection, we consider the situation when not all shocks are healable. A fixed proportion $p$ of shocks never heal; that is, their damage is permanent. Equivalently, for such shocks $\tau=0$. We incorporate the effect of such shocks, not by an increase in accumulated damage, but by a sudden drop in the threshold boundary. Figure \ref{system_subcase2} illustrates the cumulative damage until it exceeds the boundary threshold.

\begin{enumerate}[label={(\arabic*)}, align=left]
    \item Classify a shock as nonhealable with probability $p$.
    \item Let $N(t)$ denote the number of shocks that have arrived by time $t$ (observed at increments of $\Delta$). Let $H_i$ be an indicator function that takes value 1 if the $i$-th shock is healable, and 0 otherwise. Then the boundary curve drops by the corresponding magnitude of the nonhealable shock, making the modified boundary
\begin{equation}
 B(t) = a + b t - c t^2 - \sum_{i=1}^{N(t)} (1-H_i) Y_i 
\end{equation}
\item We record the cumulative damage inflicted by healable shocks only.
Therefore, the cumulative damage $D$ to the system at time $t$, is calculated as in Subsection \ref{subsec1}
\begin{equation}
 D(t) = \sum_{i=1}^{N(t)}H_i~Y_i e^{-\kappa[(t-S_i)\wedge\tau]} 
\end{equation}
\end{enumerate}

\begin{table}[ht]
\centering
\caption {For $p=0.2$ proportion of all shocks nonhealable, for $X \sim$ Weibull(2,$2/\pi$), $Y \sim$ Weibull(2,1/2), and for various choices of $\kappa$ and $B(t)$, the optimal $d$ and [the associated $t^*_{\gamma}$] are displayed for $\gamma=0.90, 0.85, 0.80$. \\}
\begin{tabularx}{0.8\textwidth} {
  | >{\raggedright\arraybackslash}X 
  | >{\centering\arraybackslash}X 
  | >{\centering\arraybackslash}X 
  | >{\centering\arraybackslash}X | }
  \hline
   \multicolumn{4}{|c|}{$B(t) = 500- t^2/60$} \\
 \hline
 $\kappa$ & $\gamma = 0.90$ & $\gamma = 0.85$ & $\gamma = 0.80$\\
 \hline
 $0.01$   & $~10.0 ~[0.263]$   &$~10.2 ~[0.367]$&   $~10.2 ~[0.476]$\\
 \hline
 $0.02$   & $~9.0 ~[0.462]$   &$~9.0 ~[0.613]$&   $~9.0 ~[0.754] $\\
 \hline
  \hline
  \multicolumn{4}{|c|}{$B(t) = 500- t^2/50$} \\
  \hline
  $\kappa$ & $\gamma = 0.90$ & $\gamma = 0.85$ & $\gamma = 0.80$\\
  \hline
  $0.01$   & $10.6 ~[0.263]$   &$10.6 ~[0.378]$&   $10.9 ~[0.485]$\\
 \hline
 $0.02$   & $~9.0 ~[0.425]$   &$9.0 ~[0.598]$&   $9.0~[0.745]$\\
  \hline
  \hline
 \multicolumn{4}{|c|}{$B(t) = 500- t^2/40$} \\
 \hline
 $\kappa$ & $\gamma = 0.90$ & $\gamma = 0.85$ & $\gamma = 0.80$\\
  \hline
  $0.01$   & $11.2 ~[0.263]$   &$11.2 ~[0.374]$&   $11.4 ~[0.480]$\\
 \hline
  $0.02$   & ~$10.3 ~[0.398]$   &$10.4 ~[0.545]$&   $10.4 ~[0.676]$\\
 \hline
 \hline
 \end{tabularx}  
\label{table_optd_subcase2}
\end{table}

As in Subsection \ref{tau_only}, here also on average, compared to Subsection \ref{simulation_original}, the overall waiting time until replacement after the alarm goes off is shorter.  Here, we do not provide any policy diagram to avoid repetition since the results are already provided in Table \ref{table_optd_subcase2}.

\subsection{Case~3: The arrival times of healable and nonhealable shocks have different distributions, so do their magnitudes} \label{types_of_shock}
Suppose that the shocks affecting the system are of two types based on their healing capabilities. The first type of shock is self-healable for a finite duration $\tau$ (or up to a certain percentage of the damage heals and the rest is permanent). We assume such healable shocks arrive with inter-arrival times $X_1, X_2,\dots, X_n$ which are IID with arbitrary CDF $F$. Moreover, the magnitudes of each of such shocks are denoted by $Y_1, Y_2,\dots, Y_n$ which are IID with arbitrary CDF $G$.
The second type of shock are nonhealable (or $\tau=0$) and their impact is characterized by drops in the non-increasing boundary threshold causing the system to degrade more severely than under natural aging. Let $Z_1, Z_2,\dots, Z_m$ denote the inter-arrival times of the nonhealable shocks, which are IID with arbitrary CDF $H$. Let $U_1, U_2,\dots, U_m$ denote the magnitudes of such shocks, which are IID with arbitrary CDF $K$.
% For the purpose of demonstrating the stochastic pathways, we assume the boundary is a quadratically decreasing function of time which experiences sporadic drops due to the arrival of non heal-able shocks. 
The system fails in one of three ways:
\begin{enumerate}[label={(\roman*)}, align=left]
    \item a new healable shock arrives so that the cumulative damage exceeds the boundary;
     \item the accumulated damage curve, although decreasing because of healing, crosses the boundary which decreases faster due to aging;
    \item a new nonhealable shock arrives so that the boundary suddenly drops below the (otherwise) gently decreasing cumulative damage curve.
   
\end{enumerate}

\begin{figure}[ht]
    \centering
    \includegraphics [scale=0.50]{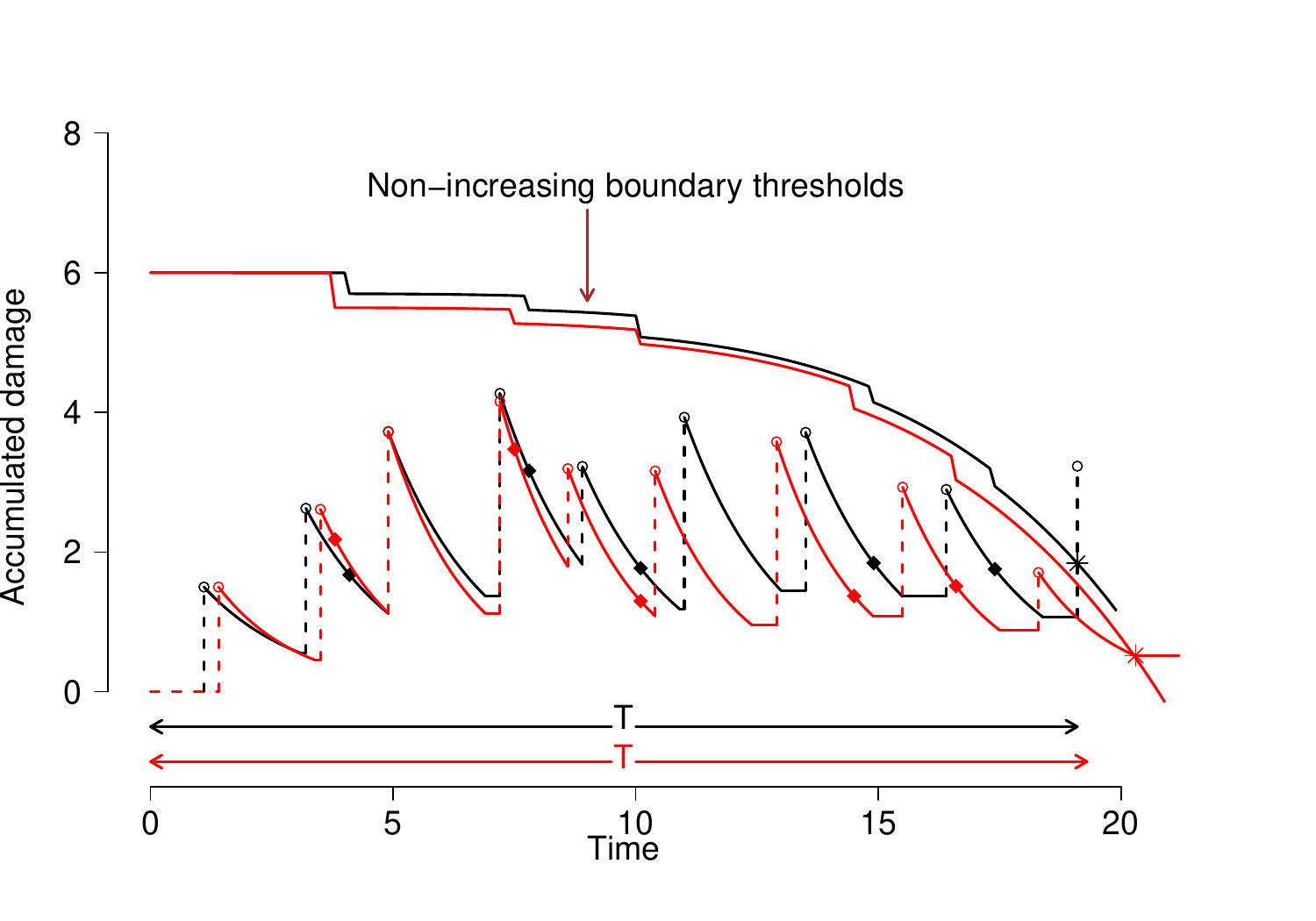}
    \caption{Depicting cumulative damage as shocks arrive randomly. The black and red stepwise decreasing curves represent the boundary threshold corresponding to the black and red sample paths respectively. Diamond-shaped dots represent the arrival times of nonhealable shocks. For illustration we consider $\tau=2$ and that nonhealable shocks arrive twice as fast as healable shocks.}
    \label{system_subcase3}
\end{figure}

Figure \ref{system_subcase3} 
illustrates the type (i) and (ii) failures using black and red sample paths which depict accumulated damage as a function of time. The black sample path crosses the boundary threshold when a healable shock of sufficient magnitude arrives. The red sample path exceeds the boundary while the system is healing, but the boundary comes down faster due to aging. Type (iii) failure is self-explanatory (and not shown in the illustration).

\begin{figure}[ht]
    \centering
   \includegraphics[width=0.70\textwidth]{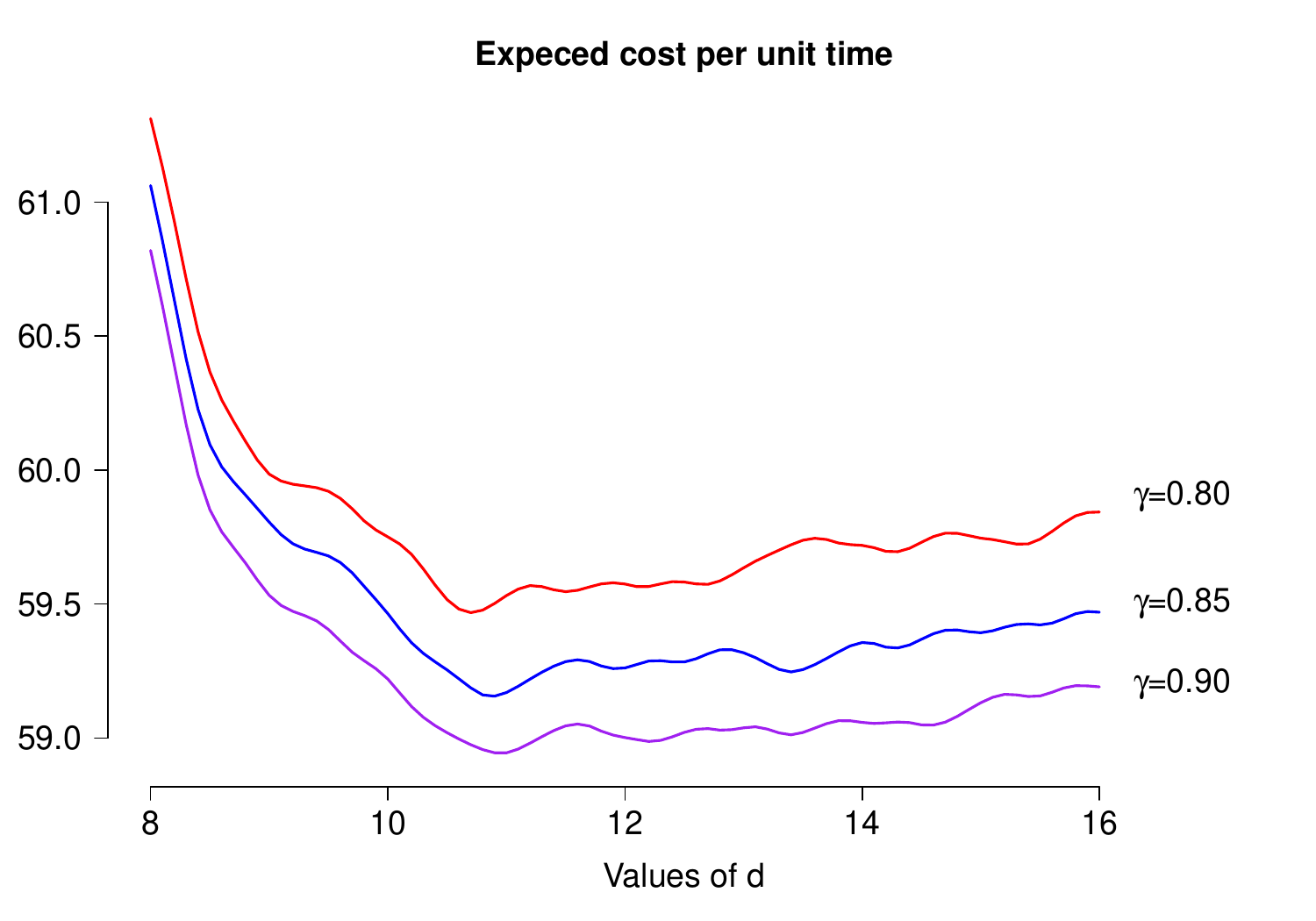}%
    \caption{The expected cost per unit time is minimized at $d = 11.1$ for $\gamma = 0.90$; at $d = 11.1$ for $\gamma = 0.85$ and $d=11.1$ for $\gamma = 0.80$. If there are multiple minima, choose the smallest one.}
    \label{policies_subcase3}
\end{figure}

Figure \ref{policies_subcase3} shows the expected cost per unit time as a function of $d$ when $B(t) = 500 - {t^2}/{50}$.
Table \ref{table_optd_subcase3} displays the simulation results showing the optimal $d$ for different choices of $\kappa$ and boundary thresholds with different quadratic coefficients, but keeping the inter-arrival time distribution of healable shocks and their magnitudes such that their means are $1$ and $10$ units respectively; and the inter-arrival times of non-healable shocks and their damage contributions such that their means are $5$ and $3$ units respectively. The given choice is considered to ensure that non-healable shocks are not more frequent than healable shocks. {Results on system reliability in subsections \ref{tau_only}, \ref{nonhealable_only}, and \ref{types_of_shock} are similar to those in Section \ref{sec2}. Hence, we do not display any new figures.}

\begin{table}[ht]
\centering
\caption {For  $X \sim$ Weibull(2,$2/\pi$), $Y \sim$ Weibull(2,1/2), $Z \sim$ Weibull(2,$10/\sqrt{\pi}$),  $U \sim$ gamma (3,1), and for various choices of $\kappa$ and $B(t)$, the optimal $d$ and [the associated $t^*_{\gamma}$] are displayed for $\gamma=0.90, 0.85, 0.80$. \\} 
\begin{tabularx}{0.8\textwidth} {
  | >{\raggedright\arraybackslash}X 
  | >{\centering\arraybackslash}X 
  | >{\centering\arraybackslash}X 
  | >{\centering\arraybackslash}X | }
  \hline
   \multicolumn{4}{|c|}{$B(t) = 500- t^2/60$} \\
 \hline
 $\kappa$ & $\gamma = 0.90$ & $\gamma = 0.85$ & $\gamma = 0.80$\\
 \hline
 $0.01$   & $~11.0 ~[0.305]$   &$~11.0 ~[0.401]$&   $~11.3 ~[0.495]$\\
 \hline
 $0.02$   & $~10.8 ~[0.415]$   &$~11.0 ~[0.543]$&   $~11.1 ~[0.647] $\\
 \hline
  \hline
  \multicolumn{4}{|c|}{$B(t) = 500- t^2/50$} \\
  \hline
  $\kappa$ & $\gamma = 0.90$ & $\gamma = 0.85$ & $\gamma = 0.80$\\
  \hline
  $0.01$   & $11.3 ~[0.290]$   &$11.3 ~[0.383]$&   $11.3 ~[0.467]$\\
 \hline
 $0.02$   & $~11.1 ~[0.382]$   &$11.1 ~[0.512]$&   $11.1 ~[0.631]$\\
  \hline
  \hline
 \multicolumn{4}{|c|}{$B(t) = 500- t^2/40$} \\
 \hline
 $\kappa$ & $\gamma = 0.90$ & $\gamma = 0.85$ & $\gamma = 0.80$\\
  \hline
  $0.01$   & $11.5 ~[0.278]$   &$11.5 ~[0.365]$&   $11.5 ~[0.448]$\\
 \hline
  $0.02$   & ~$11.1 ~[0.382]$   &$11.5 ~[0.502]$&   $12.4 ~[0.605]$\\
 \hline
 \hline
 \end{tabularx}
\label{table_optd_subcase3}
\end{table}

\section{Summary and Discussion}\label{sec4}

In this paper, we let external shocks inflict damage of varying magnitudes. We also allow the system to begin to heal instantaneously and continue to self-heal exponentially at a fixed rate $\kappa~(>0)$ while also continually degrading due to aging. We have designed a time-dependent maintenance policy that focuses on risk assessment of the system at a given time as soon as the system comes dangerously close (that is, within $d$ units) to the boundary threshold, at which instant we are warned of a high probability of failure in the near future, and thus we determine an optimal $d$ and associated replacement time by minimizing the cost per unit time. Our study allows (i) changes in healing behavior such as healing happening only for a fixed duration $\tau$; (ii) changes in types of shocks, wherein with a certain probability $p$, some shocks are healable and the others are non-healable which leave some permanent damage to the system by suddenly degrading the system by a random amount; and also a combination of both types of shocks. Here too we allow arbitrary inter-arrival time distribution of all types of shocks. If not exactly, most of the situations mentioned in this research can be found in MEMS.
We make the following important discoveries from this research that would enable the system manager to make appropriate decisions: 
 \begin{itemize}
 
 \item As the boundary degrades faster, the optimal $d$ increases, and the associated expected $t_{\gamma}^*(d)$ decreases, which means due to higher risk, we allow the system to not come too close to the boundary threshold and also allow it to run for a shorter duration of time once the risk is detected; also when the healing rate becomes faster, we can allow the optimal $d$ to be smaller and thus allow the system to function for a little bit longer. 

 \item When shocks do not heal indefinitely, but rather for a fixed duration $\tau$, their residual damage brings the collective damage closer to the boundary threshold much earlier. In general, we also see that the $t_\gamma^*(d)$'s are significantly lower and the optimal $d$'s are larger as compared to the former setup implying that the alarm goes off when the distance from the boundary is larger and we wait for a shorter duration after the alarm goes off to replace the system. Furthermore, if $\tau$ is shorter, the optimal $d$' s are even larger and the expected $t_\gamma^*(d)$'s are even smaller because healing continues for a shorter duration, increasing the risk of failure. 
 \item When a fixed proportion $p$ of shocks never heal; that is, their damage is permanent, we see that on average, as compared to Subsection \ref{simulation_original}, the overall waiting time until replacement after the alarm goes off is shorter. 
 \item When there are two types of shock where the first type of shock is self-healable for a finite duration $\tau$ (or when a certain percentage of the damage heals and the rest is permanent), and the second type of shock is non-healable, we have similar interpretations: As healing patterns change or the system degrades much faster, we wait a relatively shorter duration of time before replacing the system to reduce the risk of failure.
 \end{itemize}
This research provides a comprehensive view of different types of shocks and degradation rates. Although the simulations and illustrations consider some standard parametric distributions, the approach can easily be replicated for any distribution where parameters can be approximated from the data. 

The various choices of inter-arrival time and magnitude of damage distributions considered in this research are solely to illustrate how to find optimal replacement times. We acknowledge that a real system may not satisfy all the assumptions that we have mentioned in Section \ref{sec2}. Therefore, the practitioner should be prepared to modify these assumptions as appropriate and imitate the probabilistic approach of this research to determine a suitable maintenance policy. 
In the future, we like to incorporate different self-healing patterns and different degradation processes such as the gamma process, Wiener process, etc. 

In summary, our research provides a computational technique to construct a reasonable preventive maintenance policy for the utmost utilization of the system lifetime 
{while allowing only a small risk of failure. By numerically determining the replacement time of the system, this paper enables the system engineer to make an optimal industrial decision. This research also demonstrates that measuring the reliability of the system alone is not optimal to make a good decision. Sometimes, a cost-based approach is more beneficial. Whereas we illustrated decision-making for one particular choice of boundary function, several choices of arrival time distribution, and some variations on the healing pattern, by generalizing such choices, the computational technique applies to many other systems. Therefore, the methodology exhibited in this research contributes to a broader study of reliability and sustainability.}

\section*{Acknowledgment}
We thank our colleagues for some discussions and feedback. 

%  \section*{Funding}
% This research did not receive any specific grant from funding agencies in the public, commercial, or
%  not-for-profit sectors.
\section*{Declaration of interests statement}
The authors declare that they have no known competing financial interests or personal relationships that could have appeared to influence the work reported in this paper.

%Bibliography
\bibliographystyle{unsrt}  
\bibliography{references}

\end{document}